\pdfoutput=1  %arXiv addition for pdflatex, permits no file extensions.
\documentclass[]{article}  
\usepackage{url,float}
\usepackage{graphicx}
\usepackage{amsmath}
\usepackage{amsfonts}
\usepackage{amssymb}
\usepackage{latexsym}

\newcommand{\hide}[1]{}

\newcommand{\ABox}{
\raisebox{3pt}{\framebox[6pt]{\rule{6pt}{0pt}}}
}
\newenvironment{proof}{{\bf Proof:}}{\hfill\ABox}

\newtheorem{theorem}{{\bf Theorem}}

\newtheorem{lemma}{Lemma}
\newcommand{\lemlab}[1]{\label{lemma:#1}}
\newcommand{\thmlab}[1]{\label{thm:#1}}

\newcommand{\figlab}[1]{\label{fig:#1}}
\newcommand{\seclab}[1]{\label{sec:#1}}
\newcommand{\secref}[1]{\ref{sec:#1}}
\newcommand{\figref}[1]{\ref{fig:#1}}

{\makeatletter
 \gdef\xxxmark{%
   \expandafter\ifx\csname @mpargs\endcsname\relax % in minipage?
     \expandafter\ifx\csname @captype\endcsname\relax % in figure/caption?
       \marginpar{xxx}% not in a caption or minipage, can use marginpar
     \else
       xxx % notice trailing space
     \fi
   \else
     xxx % notice trailing space
   \fi}
 \gdef\xxx{\@ifnextchar[\xxx@lab\xxx@nolab}
 \long\gdef\xxx@lab[#1]#2{{\bf [\xxxmark #2 ---{\sc #1}]}}
 \long\gdef\xxx@nolab#1{{\bf [\xxxmark #1]}}
 \gdef\turnoffxxx{\long\gdef\xxx@lab[##1]##2{}\long\gdef\xxx@nolab##1{}}%
}

\def\P{{\mathcal P}}
\def\r{{\rho}}
\def\s{{\sigma}}
\def\S{{\Sigma}}

\newcommand{\squeezelist}{\setlength{\itemsep}{0pt}}
\title{%
On Flat Polyhedra \\
deriving from Alexandrov's Theorem
} %title

\author{%
Joseph O'Rourke%
    \thanks{Department of Computer Science, Smith College, Northampton, MA
      01063, USA.
      \protect\url{orourke@cs.smith.edu}.}
}%author

\begin{document}
\maketitle

\begin{abstract}
We show that there is a straightforward algorithm to
determine if the polyhedron guaranteed to exist by Alexandrov's
gluing theorem is a degenerate flat polyhedron,
and to reconstruct it from the gluing instructions.
The algorithm runs in $O(n^3)$ time for polygons whose
gluings are specified by $n$ labels.
\end{abstract}

\section{Introduction}
\seclab{Introduction}
A theorem of Alexandrov says that any ``gluing'' of polygons that
satisfies three conditions corresponds to a unique convex polyhedron.
The theorem includes flat doubly covered convex polygons as
among the possible ``convex polyhedra'' whose existence is
guaranteed by the theorem.
In this note we provide an algorithm that detects if a gluing will
produce such a flat polyhedron, and if so, constructs it.

Alexandrov's 1941 theorem is described in his 1950 book 
\emph{Convex Polyhedra}, recently translated into
English~\cite{a-cp-05}.
Descriptions may be found in~\cite[Sec.~23.3]{do-gfalop-07}
and~\cite[Sec.~37]{p-ldpg-10}.
Here we give a brief statement of the theorem.
Define an \emph{Alexandrov gluing} of a collection of polygons
one as satisfying these conditions:

\begin{enumerate}
\item The gluing matches all the perimeters of the polygons
by identifying which points \emph{glue} to which.
A case of special interest is when there is just one polygon,
whose perimeter is glued to itself.
Isolated points may have no match, where the boundary 
``zips'' closed in a neighborhood of those points.
\item The gluing creates no more than $2 \pi$ surface 
angle surrounding any point
of the resulting manifold.
\item The gluing results in manifold that is homeomorphic to a sphere.
\end{enumerate}

These three conditions are obviously necessary for the manifold to be
a convex polyhedron.  Alexandrov's Theorem says that these conditions
are also sufficient:

\begin{theorem}[Alexandrov]
Any Alexandrov gluing corresponds to a unique convex polyhedron
(where a doubly covered polygon is considered a polyhedron).
\end{theorem}

Alexandrov's proof is a difficult existence proof and gives little hint of the
structure
of the polyhedron guaranteed by the theorem.
Recently, Bobenko and Izmestiev found an intricate but constructive proof of
the theorem, which can be used to reconstruct the 3D polyhedron as the
solution of a  particular differential equation~\cite{bi-atwdt-06}.
They have implemented an approximate numerical solution of this equation
in publicly available software.
See~\cite{o-cgc49-07} for a high-level description of their proof.

The flat polyhedra permitted by Alexandrov's Theorem are
necessary.  For example, folding a square across a diagonal
constitutes an Alexandrov gluing, and results in a flat
doubly covered isosceles right triangle.
The purpose of this note is to isolate the degenerate flat-polyhedron
case of Alexandrov's Theorem, and show that detection and
reconstruction
are possible by a straightforward algorithm that need not
confront the complexities of the full theorem.

\section{What is $n$?}
\seclab{WhatIsN}
Before describing the algorithm, we first address a confusing issue\footnote{
First brought to my attention by Anna Lubiw.} 
concerning
the appropriate value of $n$ for this question, the primary combinatorial count for
expressing complexity.
There are four possible $n$'s:
\begin{enumerate}
\squeezelist
\item $n_p$: the total number of vertices in the collection of polygons.
\item $n_g$: the number of gluing labels defining the gluing instructions.
\item $n_s$: the number of vertices on the surface of the polyhedron $\P$.
\item $n_c$: the number of corners or \emph{cone points} on the surface of $\P$.
\end{enumerate}
A cone point is a point surrounded by strictly less than $2 \pi$ of surface.
Although we are only interested in order of magnitudes when claiming
a time complexity of $O(n^3)$, it is not the case that all these possible $n$'s
are necessarily linearly related.

For $n_p$, it is natural to count only vertices whose internal angle differs from $\pi$.
But $n_g$ could be arbitrarily larger than $n_p$.
An example is shown in 
Figure~\figref{RectangleSpiral} (other examples may be found in~\cite{addmru-cdsdob-11}).
%%%%%%%%%%%%%%%%%%%%%%%%%%%%%%%%%Figure Begin
\begin{figure}[htbp]
\centering
\includegraphics[width=0.75\linewidth]{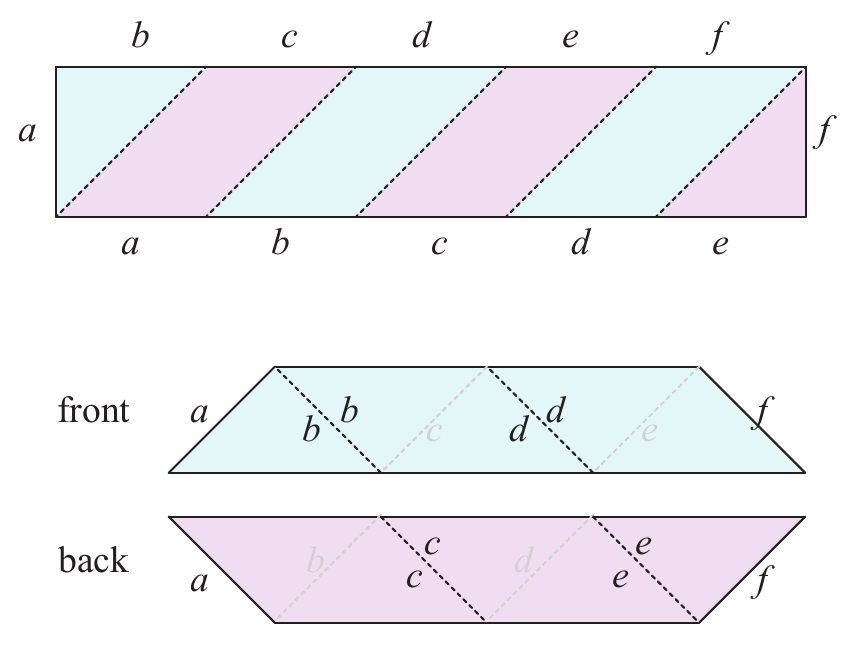}
%\fbox{X}
\caption{Top: rectangle with fold creases: $n_p{=}4$, $n_g{=}12$.
Bottom: Two views of doubly covered trapezoid: $n_s{=}7$, $n_c{=}4$.}
\figlab{RectangleSpiral}
\end{figure}
%%%%%%%%%%%%%%%%%%%%%%%%%%%%%%%%%Figure End
Here a rectangle ($n_p{=}4$) is wrapped in a spiral to form a doubly covered trapezoid ($n_c{=}4$).
But $n_g{=}12$ segments around the boundary of the rectangle are labeled to specify
the Alexandrov gluing, and it is clear that more spiraling could raise $n_g$ arbitrarily.
The gluing reduces the points on $\P$ delimiting the ``faces" by possibly half,
in this case to $n_s{=}7$. Three of these points on the trapezoid sides are not
cone points, having $2\pi$ of surface surrounding them.

For the algorithm described in the next section, it is $n_s$ that largely determines the complexity.
Because $n_s$ and $n_g$ are linearly related, and $n_g > n_s$, it is cleanest to define $n=n_g$,
the complexity of the gluing instructions.
In many cases, all four $n$'s are linearly related, but in general
it could be that $n_g$ (and so $n_s$) are arbitrarily larger than $n_p$ and $n_c$.
We must have $n_g \ge n_p$ and $n_s \ge n_c$.

The relationship between $n_p$ and $n_c$ is quite close.
Not each of the $n_p$ vertices of a polygon
necessarily ends up as a vertex of $\P$, because
vertices of the polygon whose angles sum
to $2 \pi$ can be glued together.
And 
not each of the $n_c$ cone points of $\P$ derives from
a polygon vertex, because
one can create a \emph{fold point} of angle $\pi$ at the interior
of a polygon edge.
Thus there is no exact relationship between $n_p$ and
$n_c$.
However, there can be at most four
fold-points~\cite[Lem.~25.3.1]{do-gfalop-07},
because the Gauss-Bonnet Theorem limits the total curvature of $\P$
to $4 \pi$,. So we have $n_c \le n_p+4$.

In summary, $n_g$ dominates all the others, so by the choice $n{=}n_g$,
we have all four $n$'s are $O(n_g)=O(n)$.

\section{The Algorithm}
\seclab{Algorithm}
The result of gluing the polygons together according to the gluing instructions
results in $\P$, which
could be called an \emph{abstract polyhedral surface}~\cite[Ex.~39.11]{p-ldpg-10}.
$\P$ has zero curvature everywhere except at its cone points.
Forming a data structure representing $\P$ can be accomplished in $O(n_g)=O(n)$ time.

The next step of the algorithm is to identify the $n_c$ cone points
of $\P$, which are vertices of the convex polyhedron $P$ guaranteed by
Alexandrov's Theorem.
(We are using $\P$ for the abstract surface and $P$ for the geometric polyhedron.)
Call them $v_1,\ldots,v_{n_c}$ in arbitrary order.
This step can be achieved in $O(n_c)=O(n)$ time.

The second step of the algorithm is to find the $\binom{n_c}{2}$ shortest paths on $\P$
from each $v_i$ to each $v_j$.  Call this set of shortest paths $\Sigma$.
Note that here we cannot be assured we can use an algorithm for finding shortest paths
on a convex polyhedron $P$ if that algorithm uses the 3D structure of
$P$,
because we only have available the abstract surface $\P$.
This excludes the use of the fastest known algorithm,
the $O(n \log n)$ algorithm in~\cite{ss-otasp-08}, whose first step
builds an oct-tree data structure around $P$ in 3D.
However, the Chen and Han algorithm~\cite{ch-spp-96} works entirely
intrinsic to the surface, and so can be applied to $\P$.
That algorithm assumes the surface is triangulated, and unfolds
the surface triangle-by-triangle.
Since triangulating a planar graph with $n_s$ vertices results in $O(n_s)$ triangles,
the appropriate count here is $n_s=O(n)$ by our choice of $n$ in Section~\secref{WhatIsN}.
The Chen and Han algorithm has time complexity $O(n^2)$.
Repeating this for each vertex $v_i$ as source results in $O(n^3)$
time for this step.  It is possible this brute-force approach to
computing all vertex-vertex shortest paths could be improved, but
we make no attempt here.

A key fact we use at this juncture is that every edge of a convex
polyhedron $P$
is the shortest path on $P$ between the two endpoint vertices it
connects.
This follows because any other path between those endpoints is not a
straight segment in 3D, and so is strictly longer.
Thus we know the unknown edges of $P$ are among the $O(n^2)$ 
shortest paths $\S$ on $\P$.

If indeed $P$ is a flat polyhedron, then it is a doubly covered
convex polygon, whose \emph{rim} $\r$ contains all the vertices
$v_1,\ldots,v_{n_c}$.
Moreover, the path on $\P$ that constitutes $\r$ must bisect the
angle
at each $v_i$, because the half-angle on one side is mirrored on the
other side.
So we look for such a path.
We now show that:
\begin{description}
\item[Claim~1.]
If $\r$ exists, it can be found in $O(n^3)$ time.
%and
\item[Claim~2.]
If $\r$ is found, then $P$ is uniquely identified 
as a flat polyhedron.
\end{description}

%see Figure~\figref{FlatTri}.
%%%%%%%%%%%%%%%%%%%%%%%%%%%%%%%%%Figure Begin
\begin{figure}[htbp]
\centering
\includegraphics[width=0.75\linewidth]{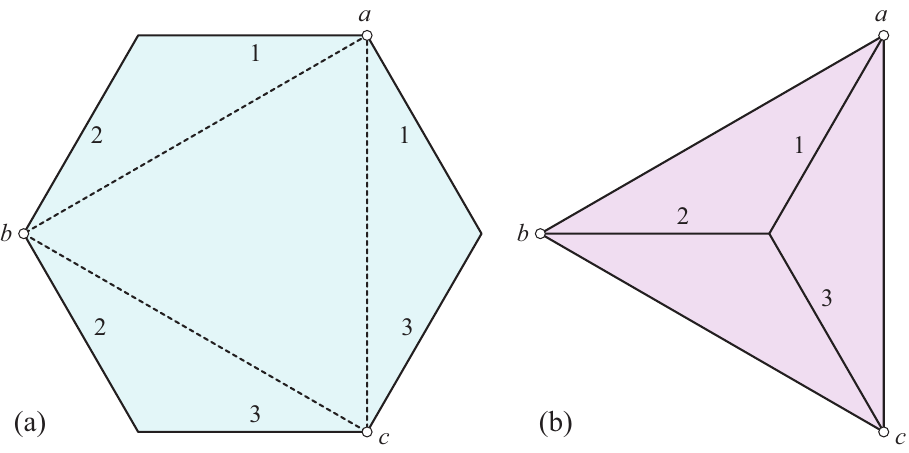}
%\fbox{FlatTri}
\caption{A regular hexagon~(a) that folds 
(toward the viewer) to a doubly covered equilateral
  triangle~(b).
The edge labels indicate gluing instructions.}
\figlab{FlatTri}
\end{figure}
%%%%%%%%%%%%%%%%%%%%%%%%%%%%%%%%%Figure End

Start with $v_1$, and look at the shortest path $\s(v_1,v_j)$
to each $v_j$, $j > 1$ in turn.
For each of these, see if it can be extended by $\s(v_j,v_k)$ so that
$(v_1,v_j,v_k)$ bisects the total angle at $v_j$.
If so, this a potential start to $\rho$, and so this path should be followed.
The path is now entirely determined by the bisection property:
the path through each vertex must bisect the total angle there.
If bisection holds at each step, and all vertices are included into
one loop, then we have found a candidate for $\rho$.
If at any stage, an outgoing bisecting shortest path is not available
from $\S$, that search can be abandoned, as it could never produce
$\rho$.
If the bisecting path closes into a loop without passing through every
vertex, again we may discard it.
In the case that we have found a candidate for $\r$, we make
one more test, for simplicity, non-self-intersection, for $\r$ must be
a simple closed path.
Although it is unclear if there can be a bisecting
self-crossing path through every vertex, 
in the absence of a proof of non-existence,
we simply check for this condition.

Let us illustrate before proceeding with the description.
Figure~\figref{FlatTri}(a) shows a regular hexagon,
whose gluing instructions fold it to an equilateral triangle~(b).%
\footnote{This example was used by Daniel Mehkeri in a 
     \emph{Math Overflow} question 4 July 10.}
In this simple example, the bisecting path $\r=(a,b,c)$ would be found immediately.

Figure~\figref{P2FlatQuad} shows a slightly more complex example,
based on~\cite[Fig.~25.24]{do-gfalop-07}.
Note the identification of the four vertices of $\P$ in~(a) of the figure.
%see Figure~\figref{P2FlatQuad}.
%%%%%%%%%%%%%%%%%%%%%%%%%%%%%%%%%Figure Begin
\begin{figure}[htbp]
\centering
\includegraphics[width=0.95\linewidth]{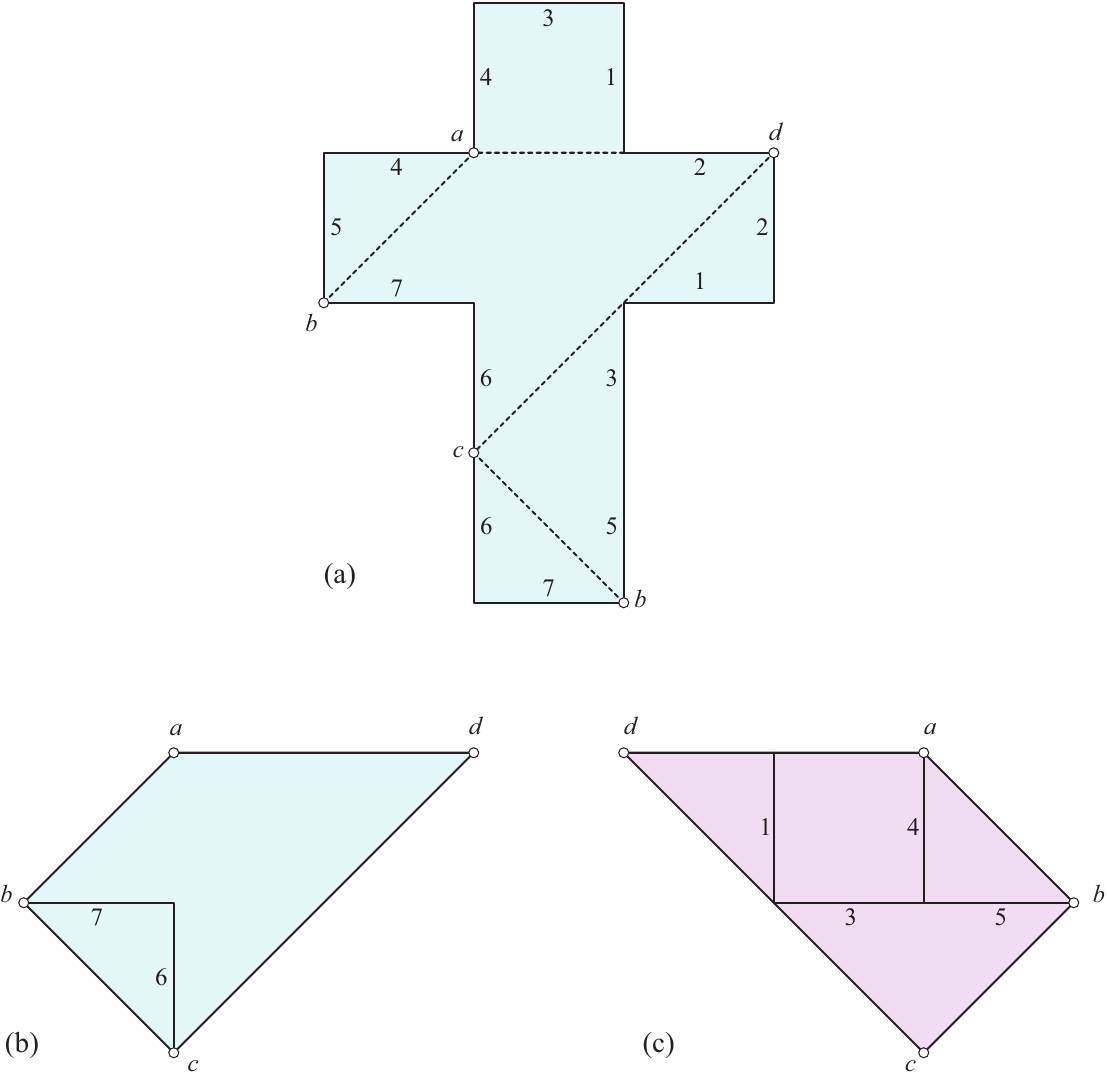}
%\fbox{P2FlatQuad}
\caption{Folding of the Latin cross~(a) 
(away from the viewer) to a doubly covered quadrilateral,
(b) and (c).
The number labels indicate gluing instruction.
}
\figlab{P2FlatQuad}
\end{figure}
%%%%%%%%%%%%%%%%%%%%%%%%%%%%%%%%%Figure End
Suppose $v_1=a$ and $v_j=c$.
The path $(a,c)$ can indeed be extended through $c$ to bisect the
angle of $\pi$ there,
but only by prematurely returning to $a$ (and then it does not bisect
at $a$).  So this path would be abandoned by the algorithm.
The path beginning $(a,b)$ continues to $\r=(a,b,c,d)$.

Returning to the argument, let us
count up the worst-case complexity of
following one $(v_1,v_j)$ path, without attempting
sophisticated algorithms.
Let us assume the shortest paths in $\S$ are maintained in sorted
order
around each source vertex, which is easily returned by the Chen and
Han algorithm.
We start with $\s(v_1,v_j)$, and search in $\S$ for a bisecting extension
$\s(v_j,v_k)$ in $O(\log n)$ time.  This same cost is incurred at
each successive step.  We also must check 
at each step if we have prematurely
closed a loop, which can be accomplished with a constant-time array
lookup of the previously visited vertices.
So a full path $\r=(v_1,v_j,\ldots,v_1)$ can be found in $O(n \log n)$ time.
Finally, we need to check $\r$ for simplicity.  Although it seems
possible
this could be accomplished in $O(n \log n)$ time, or even in $O(n)$
(because determining whether a polygon is simple 
can be accomplished in $O(n)$ time~\cite{c-tsplt-91a}),
let us just count this as $O(n^2)$ by a brute-force comparison of
every
pair of edges.

We repeat this procedure for the $n-1$ possible starts $(v_1, v_j)$,
and so spend $O(n^3)$ time overall either finding a $\r$, or determining
that
no such $\r$ exists.  If there is such a $\r$, we
must find it by this procedure, because it must pass through $v_1$.
So we have established Claim~1 above.

Claim~2 is that if we find $\r$, then indeed $P$ must be the doubly
covered flat convex polygon whose boundary is $\r$.
Here we can employ this result from~\cite[Cor.~4]{iov-sucpql-10}:

\begin{lemma}
A convex polyhedral manifold with convex boundary and with no interior
curvature
is isometric to a planar convex polygon.
\end{lemma}

\noindent
The proof of this lemma uses both Alexandrov's Theorem and a 
separate lemma of Alexandrov.
Because $\r$ includes all vertices, the interior is indeed curvature-free,
so each ``half'' of $\P$ bounded by $\r$ is isometric to a planar
convex polygon.
So we know we have a doubly covered convex polygon $P$.
Finally, Alexandrov's Theorem establishes that $P$ is unique, so there
is no need to seek another $\r$.

\section{Conclusion}
\seclab{Conclusion}
Although reconstructing the 3D structure of the polyhedron guaranteed to
exist
by Alexandrov's Theorem is a challenging problem, it is relatively
easy
to detect the degenerate flat-polyhedron case of the theorem,
and to reconstruct the doubly covered convex polygon:
Just follow vertex-to-vertex shortest paths seeking the rim.
Although the algorithm described has cubic time complexity,
it seems possible that it could be reduced to near-quadratic
complexity.

Returning to the different $n$'s discussed in Section~\secref{WhatIsN},
the time complexity is $O(n_g)$ to form the abstract surface $\P$,
$O(n_s^2 n_c)$ to find the shortest paths, and $O(n_c^3)$ to find $\r$.
The factor that dominates is $n_s^2$, so it could be worthwhile
to reduce $n_s$ by merging coplanar faces prior to running
the shortest-paths algorithm.

Perhaps a more interesting direction for future research is to explore
whether other special classes of polyhedra $P$ might be
reconstructable from an Alexandrov gluing
without all the machinery of~\cite{bi-atwdt-06}.

\paragraph{Acknowledgments.}
The idea for this note resulted from a stimulating 
workshop conversation
with
Alexander Bobenko,
Ivan Izmestiev, and
Konrad Polthier in 2007.
I thank Anna Lubiw for raising the wrapping issue
discussed in Section~\secref{WhatIsN}.

\bibliographystyle{alpha}
\bibliography{/Users/orourke/bib/geom/geom}
\end{document}